\documentclass{article}

\usepackage{arxiv}

\usepackage[utf8]{inputenc} \usepackage[T1]{fontenc}    

\PassOptionsToPackage{hyphens}{url}

\usepackage[breaklinks,backref=page]{hyperref}

\usepackage{url}            \usepackage{booktabs}       \usepackage{amsfonts}       \usepackage{nicefrac}       \usepackage{microtype}      \usepackage{lipsum}         \usepackage{graphicx}
\usepackage[numbers,sort&compress]{natbib}
\usepackage{doi}
\usepackage{float}
\usepackage{multirow}
\usepackage{lineno}

\usepackage{cleveref}

\usepackage{wrapfig}

\usepackage{authblk}

\usepackage[justification=centering]{caption}

\usepackage[inline]{enumitem}

\usepackage{bbding}

\usepackage[flushleft]{threeparttable}

\usepackage{listings}

\usepackage{subcaption}

\renewcommand*{\backref}[1]{}
\renewcommand*{\backrefalt}[4]{
  \ifcase #1
    No citations.\or
(Cited on page: #4).
  \else
(Cited on pages: #4).
  \fi
}

\title{Esports Training in StarCraft II: Stance Stability and Grip Strength }

\date{June 15, 2024}

\makeatletter
\newcommand\email[2][]{\newaffiltrue\let\AB@blk@and\AB@pand
      \if\relax#1\relax\def\AB@note{\AB@thenote}\else\def\AB@note{\relax}
        \setcounter{Maxaffil}{0}\fi
      \begingroup
        \let\protect\@unexpandable@protect
        \def\thanks{\protect\thanks}\def\footnote{\protect\footnote}
        \@temptokena=\expandafter{\AB@authors}
        {\def\\{\protect\\\protect\Affilfont}\xdef\AB@temp{#2}}
         \xdef\AB@authors{\the\@temptokena\AB@las\AB@au@str
         \protect\\[\affilsep]\protect\Affilfont\AB@temp}
         \gdef\AB@las{}\gdef\AB@au@str{}
        {\def\\{, \ignorespaces}\xdef\AB@temp{#2}}
        \@temptokena=\expandafter{\AB@affillist}
        \xdef\AB@affillist{\the\@temptokena \AB@affilsep
          \AB@affilnote{}\protect\Affilfont\AB@temp}
      \endgroup
       \let\AB@affilsep\AB@affilsepx
}
\makeatother

\author[1]{\textbf{Andrzej Białecki}\textsuperscript{*,}} 
\affil[1]{Warsaw University of Technology}

\author[2]{\textbf{Michał Staniszewski}}
\author[2]{\textbf{Robert Białecki}}
\author[2]{\textbf{Jan Gajewski}}
\affil[2]{Józef Piłsudski Warsaw University of Physical Education}

\hypersetup{
pdftitle={Esports Training in StarCraft II: Stance Stability and Grip Strength},
pdfauthor={Andrzej Białecki},
pdfkeywords={sports, grip strength, stance stability, biomechanics, StarCraft II},
}

\providecommand{\parencite}[1]{}

\renewcommand{\parencite}[1]{\cite{#1}}

\begin{document}
\renewcommand{\figureautorefname}{Fig.}
\renewcommand{\subsectionautorefname}{Subsection}

\maketitle

\begin{abstract}

    Esports are a mostly sedentary activity. There is a growing need for investigation into how biomechanical and physical abilities can be optimized for esports through training. One such research avenue concerns the ability of esports players to perform balance tasks due to the prolonged sedentary states that are required to reach the top echelon of performance.

    Our aim for this work is to describe and compare physical abilities (balance, grip strength, and self-reported training habits) of top Polish StarCraft~2 tournament players.

    Esports players differed significantly from the reference group in their ability to balance on one leg. Additionally, in a grip strength test, the esports group fared worse than the reference group in all consecutive attempts.

    Despite self-reported physical activity in the esports group, player fitness requires further research. Training optimization could offset the issues arising from sedentary activity, and intensifying esports training so it could take less time overall.

\end{abstract}


\section{Introduction}

Training and competing in most esports is a sedentary endeavor. By extension, esports in their majority are desribed as a cognitive activity \parencite{Nicholson2024SelfReporting}. Despite that, players self-report in interviews and questionnaires that they know about the benefits of physical activity \parencite{Bialecki2024ESPORT}. Such claims can be over-estimated as shown by recent research leveraging accelerometer based data to measure activity levels \parencite{Nicholson2024SelfReporting}. And yet, other authors indicate that time spent playing video games (not in a professional sense) is unlikely to impact self-reported well-being \parencite{Vuorre2022}. On the other hand, it is well established that functional physical activity is key to life longevity and reducing the chance of sustaining injuries \parencite{Daly2008,Morena2021}. Even so, some injuries and physical abilities may not be critical to esports performance. It is clear that esports requires more research towards optimizing training through introducing physical activity - especially to minimize the loss of physical abilities sustained by a sedentary lifestyle and profession.

Psychological, physiological and biomechanical indices are widely researched and utilized in sports. For psychology this includes eye tracking, psychometric tests, reaction time and others \parencite{Rhonda2018,Mikicin2022,Koch2021,Piepiora2023}. In the case of physiology there are heart rate, lactate, creatine kinase and other relevant information \parencite{Adamczyk2023}. Finally, when looking into biomechanical features, we find positional tracking, maximal torque, and acceleration \parencite{Ingwersen2023,PachavaGoel2022}. Yet this information is not exhaustive, and there are surely more measureable data that can be used.
And yet, biomechanical indices describing players and their relation to performance in gaming and esports are not fully developed \parencite{Dupuy2024}. There are some published works containing multimodal esports data containing all of the above areas of research (psychology, physiology, biomechanics) and works that attempt to connect bio-signals of varying nature that hold promise for future research in multiple directions \parencite{Smerdov2022AISensors,Smerdov2019Chair,Smerdov2021Burnout}.

Given that research on the physiological \parencite{Nicholson2024EnergyExpenditure,Katelhut2024} and biomechanical profile of esports players is still developing \parencite{Dupuy2024,Lam2022}, we have identified a knowledge gap regarding the comparison of esports to other major sports in terms of the basic physical abilities of players.

The aim of our work is to describe and compare the physical abilities (balance, grip strength, and self-reported training habits) of top Polish StarCraft~2 tournament players and a similarly aged reference group consisting of students attending Józef Piłsudski University of Physical Education in Warsaw.

To fulfill our aim we have formulated the following research questions:
\begin{itemize}
    \item[\textbf{RQ 1:}] What are the differences between esports players and the reference group in their ability to perform stabilometric tasks?
    \item[\textbf{RQ 2:}] What is the difference between the maximum grip strength when comparing esports players and a group of physically active non-esports gamers or non-players?
    \item[\textbf{RQ 3:}] What are the characteristics of the self-reported training structure in StarCraft 2 esports?
\end{itemize}
\section{Material and Methods}
\label{sec:material_methods}

\subsection{Participants}
\label{sec:participants}

In total we measured 35 voluntary participants who were split into two groups. The first group consisted of eleven participants, including top Polish StarCraft 2 players (n=10) and one player whose main game was Counter-Strike (n=1). Measurements for this group were conducted during a community organized competition at the ESPOT Tournament in Warsaw. All of the players qualified for the tournament either by invitation or took part in multiple qualification tournaments. The second group consisted of students attending Józef Piłsudski University of Physical Education in Warsaw (n=24). In this group measurements were conducted during university classes. Descriptive statistics for the participants are presented in \autoref{tab:measurement_stats}.

\begin{table*}[!h]
    \centering
    \begin{threeparttable}
        \caption{Summary of the measured groups.}
        \label{tab:measurement_stats}
        \begin{tabular}{lcc}
            \hline
                            & reference group (n = 24)                     & esports players (n = 11)                  \\ \hline
                            & \multicolumn{2}{c}{mean $\pm$ SD (min; max)}                                             \\ \hline
            age             & 21.9 $\pm$ 1.4 (19.4; 24.3)                  & 24.9 $\pm$ 3.7 (18.8; 29.3) $^{\ast\ast}$ \\
            height {[}cm{]} & 181.5 $\pm$ 8.3 (163.0; 202.0)               & 183.9 $\pm$ 8.5 (172.0; 200.0)            \\
            weight {[}kg{]} & 77.6 $\pm$ 11.2 (58.0; 105.0)                & 79.1 $\pm$ 19.8 (60.0; 130.0)             \\
            shoe size       & 43.9 $\pm$ 1.9 (41.0; 47.5)                  & 43.6 $\pm$ 2.3 (41.0; 49.0)               \\ \hline
        \end{tabular}
        \begin{tablenotes}
            \footnotesize
            \item[\raisebox{-0.5ex}{**}] p $<$ 0.01
        \end{tablenotes}
    \end{threeparttable}
\end{table*}

\subsection{Study Design}
\label{sec:study_design}

Each of the participants was asked to fill out a form consisting of 76 questions on their approach to various training aspects, which included their number of training units during a day, physical preparation, nutrition, and others. Out of the esports players group all (n=11) participants filled out the questionnaire. Not all reference group participants filled out the questionnaire (n=13/24).

To measure grip strength a hydraulic hand dynamometer SH5001 manufactured by Saehan Corporation in Masan, Korea was used. The grip strength measurement protocol consisted of six consecutive attempts while standing with the arm parallel to the body. Each attempt was followed by switching the tested hand, starting with the left hand. Participants were instructed to try their best to exert maximum force on the measurement device.

For the balance testing our measurement leveraged a Sensor Medica FreeMed (Italy) stabilometric mat with included FreeStep software. Prior to data capture, each participant received instructions on the measurement process. The protocol for measuring participants ability to balance the software inferred center of mass was: (0) Removing the footware, standing on the mat, getting used to the new surface, (1) 30 seconds on both feet with eyes open, (2) 30 seconds on both feet with eyes closed, (3) 10 seconds on left foot with eyes open, (4) 10 seconds on left foot with eyes closed, (5) 10 seconds on right foot with eyes open, and (6) 10 seconds on right foot with eyes closed.

\subsection{Statistical Methods}
\label{sec:statistical_methods}

Statistical analyses were carried out using Statistica~13.3 (TIBCO Software Incorporation). For the initial investigation descriptive statistics were calculated and we performed an independent samples t-test for antropomorphical features. Further, while the distribution of grip strength was normal, the center of pressure (CoP) path length had to be log-transformed. We employed Shapiro-Wilk tests to verify the normality of feature distributions. Comparisons of mean values were made using analysis of variance for repeated measures (RM ANOVA).

For grip strength, a fixed factor of ``GROUP'' (student, esports player) and two repeated factors of ``MEASUREMENT'' (1, 2, 3) and ``SIDE'' (right, left) were included. For comparisons of the path length of the centre of pressure in standing with both feet, the constant factor ``GROUP'' and the repeated factor ``EYES'' (open, closed) were taken into consideration. In single-leg standing, the ``SIDE'' (right, left) factor was additionally included. A significance level of $\alpha=0.05$ was used. Effect sizes were assessed using partial eta squared ($\eta^{2}$). Post-hoc comparisons were done using the Tukey test.

\section{Results}

Initial investigation into the differences between groups did not reveal a significant difference in basic antropomorphical features. The only observed difference was in the average age of participants: 21.9 for the reference group and 24.9 for the esports group, see \autoref{tab:measurement_stats}. Such results indicate that groups were selected correctly, and predominantly differed in their sporting activities. The randomly sampled student reference group was fit to continue further investigation into other measured aspects.

\subsection{Self-Reported Training Questionnaire}

As mentioned in \nameref{sec:study_design}, participants self-reported on selected aspects of their training regime. When asked about their esports/sports experience in years, the self reported values for the groups were as follows: esports group (n=11/11, average: 10.27, SD: $\pm$3.41, min: 4, max: 17), reference group (n=13/24, average: 6.64, SD: $\pm$2.99, min: 3, max: 14).

When asked about their participation in other sporting activities outside of their main esport, esports players largely confirmed (n=9/11) that they had prior experience in other sporting activities such as: football (n=5), track and field (n=2), basketball (n=2), volleyball (n=2), swimming (n=1), chess (n=1), tennis (n=1), weightlifting (n=1), and rowing (n=1). In total 16 activities were reported.

Similarly, the reference listed various games that they play recreationally such as: League of Legends (n=4), Counter-Strike (n=3), and FIFA (n=2); nine in total.

When asked about the availability of a coach, only one player (n=1/11) from the esports group responded that they have such access. Overwhelmingly, both professional gamers and the reference group participants stated that they do not have a short-term (weekly), medium-term (monthly), or long-term (annual) training plan. Only one esports player (n=1/11) and one student (n=1/13) declared that they implement the training process according to a preconceived plan.

\begin{figure}[H]
    \centering
    \includegraphics[width=0.8\textwidth]{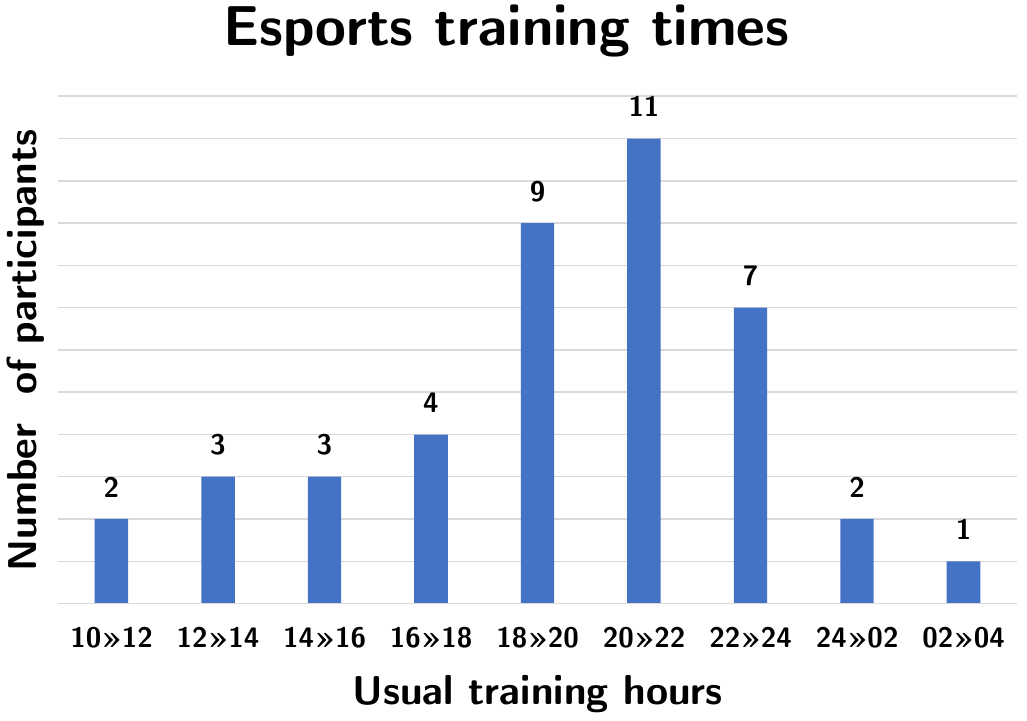}
    \caption{Self-reported most frequent training times for the esports group.}
    \label{fig:training_times}       \end{figure}

\begin{figure}[H]
    \centering
    \includegraphics[width=0.6\textwidth]{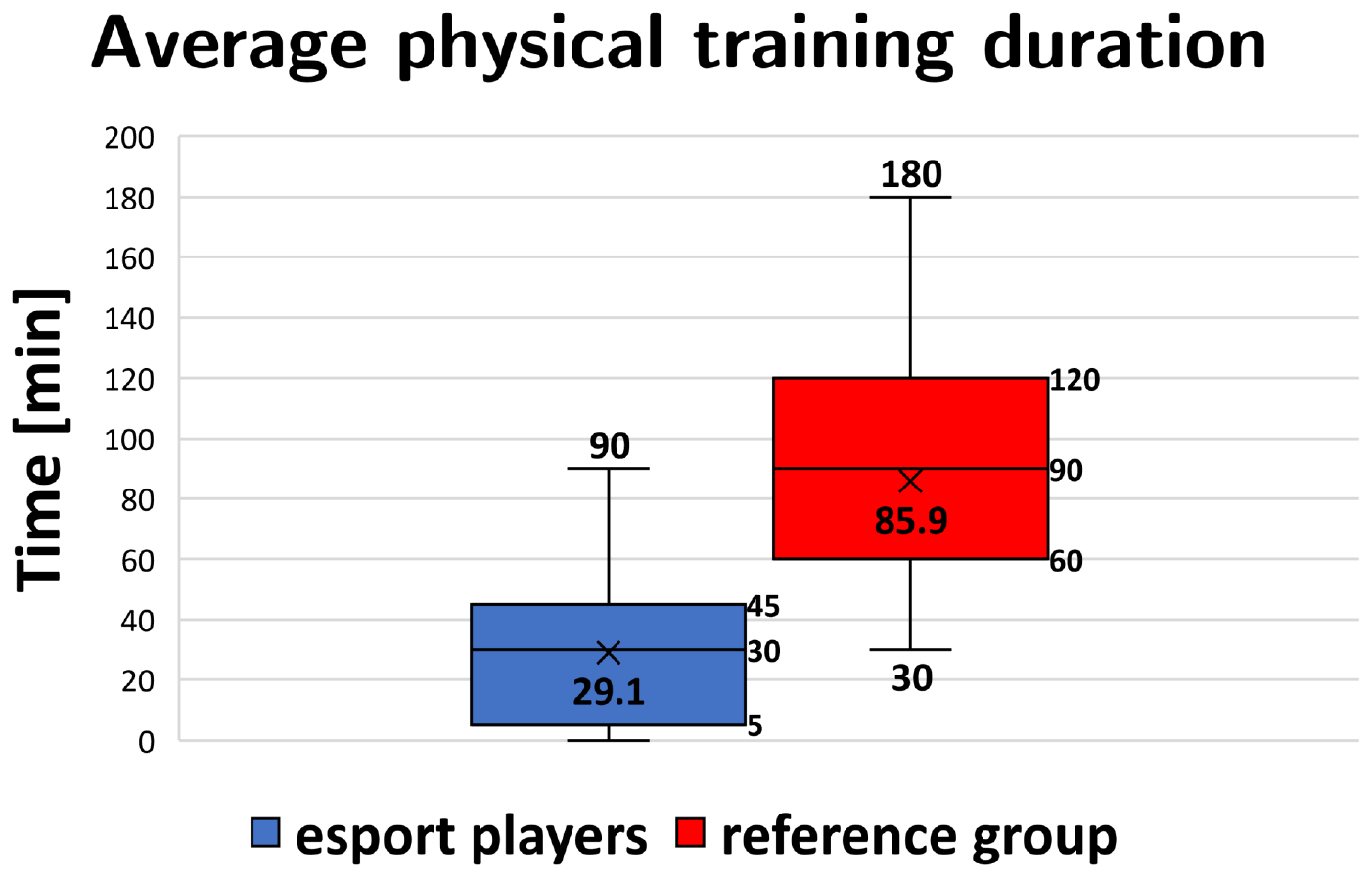}
    \caption{Self-reported physical training time in both groups.}
    \label{fig:training_times_physical}       \end{figure}

Additionally, the primary hours of training were found to be between 18.00 and 24.00, as shown on \autoref{fig:training_times}. Moreover, players reported that on average the time that they take for a single training session is as follows: n=11/11, average: 188.18 [min], SD: $\pm$101.61 [min], min: 60 [min], max: 420 [min]. Declared physical traininig time is shown in \autoref{fig:training_times_physical}

Regarding whether the esports group controls their training through a diary, only one participant confirmed such use (n=1/11). Concerning keeping track of the training/gameplay statistics in some way, only four esports players responded positively (n=4/11). When asked about their mouse grip, the esports group responded with: Palm (n=6), Fingertip (n=4), and Claw (n=1). As for hotkey customizations, the majority of esports players stated that they use their own game settings (n=10/11). All participants (n=24) self-reported that they hold the mouse in their right hand.

\subsection{Stabilometry and Grip Strength}

Results concerning balance measurements indicate that there are no confirmed significant differences between groups in the bipedal task as shown in \autoref{tab:results_both_legs}. At the same time, we observed statistically significant results indicating that the measured esports players were worse at balancing on a single leg. These results indicated an interaction between all tested factors of ``GROUP'', ``EYES'', and ``GROUP $\times$ EYES'', see \autoref{tab:results_single_leg}. Considering measurements with eyes open and closed, the CoP path length for standing on the right limb in esports players was longer than that for the left limb ($p<0.01$) and longer than that for students ($p<0.01$ and $p<0.001$ for the right and left limbs respectively).

\begin{table*}[!h]
    \centering
    \begin{threeparttable}
        \caption{CoP path length [mm] for bipedal measurement with eyes open and closed for reference group and esports players.}
        \label{tab:results_both_legs}
        \begin{tabular}{lcc}
            \hline
                   & students (n = 24)                            & esports players (n = 11) \\ \hline
            eyes   & \multicolumn{2}{c}{mean $\pm$ SD (min; max)}                            \\ \hline
            open   & 204 $\pm$ 32 (128; 266)                      & 241 $\pm$ 66 (175; 348)  \\
            closed & 229 $\pm$ 68 (168; 453)                      & 249 $\pm$ 62 (170; 355)  \\ \hline
        \end{tabular}
        \begin{tablenotes}
            \footnotesize
            \item[Effect group: $F_{1,34}=2.12$, $p=0.1542$, $\eta^{2}=0.059$]
            \item[Effect eyes: $F_{1,34}=1.85$, $p=0.1823$, $\eta^{2}=0.052$]
            \item[Effect group $\times$ eyes: $F_{1,34}=0.47$, $p=0.4976$, $\eta^{2}=0.014$]
        \end{tablenotes}
    \end{threeparttable}
\end{table*}

\begin{table*}[!h]
    \centering
    \begin{threeparttable}
        \caption{CoP path length [mm] for the single leg task with eyes open and closed.}
        \label{tab:results_single_leg}
        \begin{tabular}{llcc}
            \hline
                                    &                          & students (n = 24)                            & esports players (n = 11)   \\ \hline
            eyes                    & \multicolumn{1}{c}{side} & \multicolumn{2}{c}{mean $\pm$ SD (min; max)}                              \\ \hline
            \multirow{2}{*}{open}   & right                    & 337 $\pm$ 87 (183; 479)                      & 505 $\pm$ 181 (384; 1027)  \\
                                    & left                     & 288 $\pm$ 92 (134; 484)                      & 364 $\pm$ 106 (212; 611)   \\
            \multirow{2}{*}{closed} & right                    & 773 $\pm$ 240 (414; 1533)                    & 1016 $\pm$ 346 (615; 1716) \\
                                    & left                     & 762 $\pm$ 259 (432; 1775)                    & 800 $\pm$ 227 (417; 1216)  \\ \hline
        \end{tabular}
        \begin{tablenotes}
            \footnotesize
            \item[Effect group: $F_{1,34}=10.77$, $p=0.0024$, $\eta^{2}=0.241$]
            \item[Effect eyes: $F_{1,34}=19.54$, $p=0.0001$, $\eta^{2}=0.365$]
            \item[Effect group $\times$ eyes: $F_{1,34}=4.93$, $p=0.0332$, $\eta^{2}=0.127$]
        \end{tablenotes}
    \end{threeparttable}
\end{table*}

\begin{figure}[thb]
    \centering
    \includegraphics[width=0.6\textwidth]{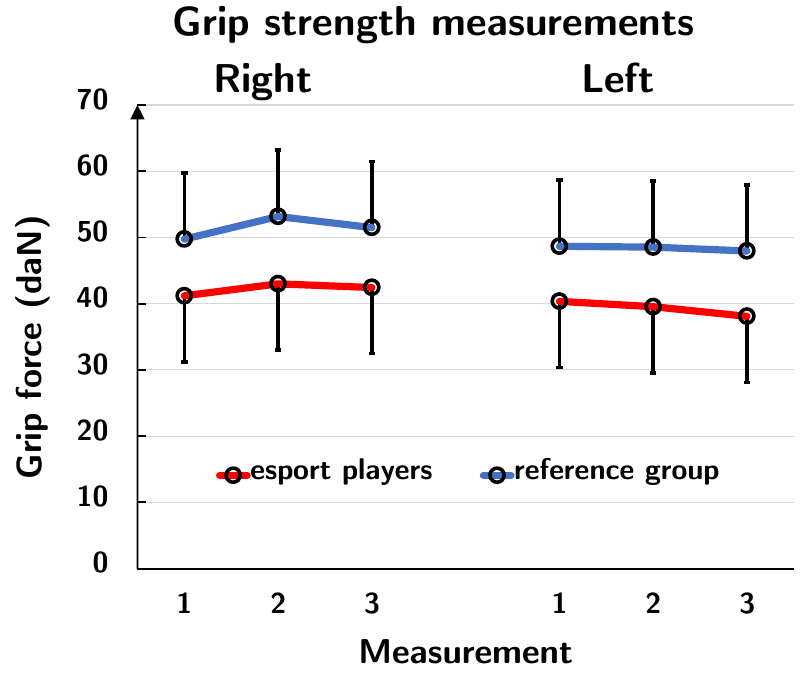}
    \caption{Results of the grip strength measurements split by group.}
    \label{fig:grip_strength_results}       \end{figure}

Analysis concerning the grip strength showed significant differences between groups and repetitions (within consecutive same hand measurements), and side; see \autoref{fig:grip_strength_results}. For the reference group grip strength was greater than that of esports players ($F_{1,33}=10.99$, $p=0.0023$, $\eta^{2}=0.25$). In both participant groups, the strength of the right hand was greater than that of the left hand ($F_{1,33}=14.62$, $p=0.0001$, $\eta^{2} = 0.31$). Differences in subsequent measurements were dependent on side. While for the right hand measurement the second try was the strongest, each consecutive try for the left hand was weaker ($F_{2,66}=5.71$, $p=0.0052$, $\eta^{2}=0.15$).

\section{Limitations}

The esports players were asked to perform the tasks in between tournament games. Due to this we cannot rule out that their ability was trumped by the specific cognitive requirements of an esports tournament. This could also include an emotional response depending on the in-game factors. Additionally, each of the participants was asked to perform the task at a different time during the day. Because of the tournament requirements we were not able to accomodate for perfect laboratory conditions.

Finally, some of the players and the reference group participants could have been biased in their self-reported responses when filling out the questionnaire and stating their weight or height, despite there being no incentive towards such bias.

Despite all of the abovementioned limitiations we feel like this work highlights some key information on the physical parameters of top StarCraft 2 Polish players. We recommend further research into incorporating physical activity in a way that can assists life longevty and esports performance.
\section{Discussion}

Based on the results of our analyses we have formulated the following responses in relation to our research questions:

\textbf{RQ 1:} Despite similar self-reported approaches to physical activity, esports players were worse at stabilometric tasks performed on a single leg than the reference group. This suggests that despite the reported physical activity, the sedentary time affected this group negatively in general.

\textbf{RQ 2:} Esports players performed worse in the maximal grip strength task than the reference group. Such results may suggest that esports players - despite the heavy use of their hands - do not require strength for in-game performance. Additionally, as StarCraft 2 requires from 200 to 300 or more actions per minute (APM), players' hands ought to be agile, coordinated, fast and flexible, but not necessarily strong.

\textbf{RQ 3:} In the case of StarCraft 2, the in-game and out-of-game training characteristics require a lot of time investment into training. Further research should be aimed at finding optimal training regimes and training tools while keeping the same level of perceived fun and accomplishment. Such progress should also consider the diversity and aim of stimuli within the training regimes so that esports athletes can receive full care despite not having access to coaches, analysts, and other infrastructure.

Standing on both feet is a natural human activity that does not require additional or specific exercises to maintain balance effectively. Therefore, it is not surprising that there are no significant differences in body stability between esports players and the reference group (physical education students). However, when standing on one foot - which requires specific skills - there were statistically significant differences between the two groups. The weaker stability among esports players is likely due to their required sedentary activity within esports and other areas of life, compared to the reference group engaging in various physical activities that stimulate their neuromuscular system. Therefore, the result may be a more stable position when standing on one foot. Similar findings were reported in traditional sports showing that intensive nine-day ski training improved stability parameters but only in measurements performed in ski boots - which is a specific activity for humans - but not when naturally standing barefoot \parencite{Staniszewski2016}. In the same way in a study of snowboarders, it was found that after nine days of intensive training there were no differences in stability parameters when standing barefoot straight (natural activity). However, a significant improvement in these parameters was observed when standing barefoot in the snowboard basic position, because riding in this position was trained (specific activity) \parencite{Staniszewski2017}. As \cite{Schorderet2021} showed in their meta-analysis, the length of the CoP pathway during single-leg standing did not differentiate between the dominant and non-dominant limb, which is in contrast to the results obtained in our study.

Similarly, stabilometric research on older adults suggests that less than 30 minutes of moderate to vigorous exercise per day and more than 11 hours of low energy expenditure as measured with CoP displacement data are typical for aging, diverse pathological conditions, and such individuals may be prone to falls. Yet, it is hard to tie such results directly to the varying physical activity levels when only mean velocity of CoP was found to be significantly different between the two groups: one that consisted of older adults sufficiently active and without sedentary behavior and the other with older adults insufficiently active and with sedentary behavior \parencite{Piropo2021}.

Other stabilometric research showcased the alleviating effects of light intensity treadmill use (two hours) on the postural sway as measured in stabilometric tasks similar to ours. This work was in favor of the workplace treadmill desk use \parencite{Charalambous2019}. Such regimes and interventions could make their way into esports training. Additionally, static and dynamic balance performance is subject to change depending on the time of measurement (diurnal fluctuation) \parencite{Dana2021}. We haven't taken into account the potential variability that could have occurred due to this factor.

\cite{Filipowicz2020} observed that an adolescent group of non-training girls exhibited worse stabilometric parameters as compared to non-training boys and training girls. In the case of our measurement the reference group could be described as physically active due to the demands of the sports university classes. We could not state the same for the group of esports players for which the true extent of physical activity was unknown.

Generally, the negative stereotypes towards video gaming and esports have not held up in recent research by \cite{Ketelhut2023Fitness}. Observations stemming from this work demonstrated that esports players are a heavily heterogenous group that spans throughout the spectrum of fitness and health features. There the comparison group was not found to be statistically different from the esports players in a maximal grip strength test.

Additionally, our results align with those of \cite{Darwesh2023} where a group of experienced video game players was found to have a statistically significant weaker grip strength as compared to a group of students that did not partake in as much of video game play. Similarly, reported by \cite{Uluagac2023}, when comparing esports players with more and less experience a statistically significant difference was observed for the dominant hand grip strength, and other related features. Upper extremity health cannot be overlooked in case of esports players as it could affect their future esports careers. Investigating biomechanical aspects related to grip strength, including neuromuscular fatigue and others, could be beneficial. Especially in light of research presented by \cite{Lee2024} diving deeper into the input precision and behavioral metrics of expert League of Legends players.

\section{Conclusions}

Based on our findings, we conclude that esports players should engage with physical activity interventions. Due to the long hours spent in a sedentary position, esports training has great potential to harm players in the long-term. Moreover, we state that esports still requires major research efforts toward describing the physical capabilities of players, and the place of physical training in esports periodization. Leveraging existing research in sports sciences, psychology, and informatics should allow for esports to become one of the best researched group of sports in the world thanks to the availability of data from game engines, and the potential for sensor fusion. That being said, we recommend further attempts to link physical capabilities to in-game performance in order to start encompassing a comprehensive model for top-level esports performance.

\pagebreak

\section*{Authors Contributions}

\begin{enumerate}
    \item Conceptualization: Andrzej Białecki;
    \item Supervision: Andrzej Białecki, Jan Gajewski;
    \item Methodology: Andrzej Białecki, Michał Staniszewski, Jan Gajewski;
    \item Formal Analysis: Andrzej Białecki, Jan Gajewski;
    \item Investigation: Andrzej Białecki, Robert Białecki, Jan Gajewski;
    \item Writing - original draft: Andrzej Białecki;
    \item Writing - review and editing: Andrzej Białecki, Michał Staniszewski, Robert Białecki, Jan Gajewski;
    \item Equipment - Michał Staniszewski, Jan Gajewski.
\end{enumerate}

\section*{Acknowledgements}

Special thanks to Paweł Dobrowolski for his ongoing support of gaming and esports research initiatives. Extended to Jakub Infulecki and Piotr Kaproń for assistance in measurements.

Additionally, we would like to thank the esports community for sharing their experiences, playing together and discussing key aspects of the gameplay in various esports. We extend our thanks especially to: Mikołaj ``Elazer'' Ogonowski, Mateusz ``Gerald'' Budziak, Igor ``Indy'' Kaczmarek, Jakub ``Trifax'' Kosior, Michał ``PAPI'' Królikowski, and Konrad ``Angry'' Pieszak.

\section*{Conflict of Interest}

Authors state that there is no conflict of interest.

\bibliographystyle{IEEEtran}
\bibliography{IEEEabrv,sources.bib}

\begin{thebibliography}{10}
\providecommand{\url}[1]{#1}
\csname url@samestyle\endcsname
\providecommand{\newblock}{\relax}
\providecommand{\bibinfo}[2]{#2}
\providecommand{\BIBentrySTDinterwordspacing}{\spaceskip=0pt\relax}
\providecommand{\BIBentryALTinterwordstretchfactor}{4}
\providecommand{\BIBentryALTinterwordspacing}{\spaceskip=\fontdimen2\font plus
\BIBentryALTinterwordstretchfactor\fontdimen3\font minus
  \fontdimen4\font\relax}
\providecommand{\BIBforeignlanguage}[2]{{%
\expandafter\ifx\csname l@#1\endcsname\relax
\typeout{** WARNING: IEEEtran.bst: No hyphenation pattern has been}%
\typeout{** loaded for the language `#1'. Using the pattern for}%
\typeout{** the default language instead.}%
\else
\language=\csname l@#1\endcsname
\fi
#2}}
\providecommand{\BIBdecl}{\relax}
\BIBdecl

\bibitem{Nicholson2024SelfReporting}
\BIBentryALTinterwordspacing
M.~Nicholson, C.~Thompson, D.~Poulus, T.~Pavey, R.~Robergs, V.~Kelly, and
  C.~McNulty, ``{Physical Activity and Self-Determination towards Exercise
  among Esports Athletes},'' \emph{Sports Medicine - Open}, vol.~10, no.~1,
  p.~40, 04 2024. [Online]. Available:
  \url{https://doi.org/10.1186/s40798-024-00700-0}
\BIBentrySTDinterwordspacing

\bibitem{Bialecki2024ESPORT}
A.~Białecki, P.~Xenopoulos, P.~Dobrowolski, R.~Białecki, and J.~Gajewski,
  ``{ESPORT: Electronic Sports Professionals Observations and Reflections on
  Training},'' 2023.

\bibitem{Vuorre2022}
\BIBentryALTinterwordspacing
M.~Vuorre, N.~Johannes, K.~Magnusson, and A.~K. Przybylski, ``{Time spent
  playing video games is unlikely to impact well-being},'' \emph{Royal Society
  Open Science}, vol.~9, no.~7, p. 220411, 2022. [Online]. Available:
  \url{https://royalsocietypublishing.org/doi/abs/10.1098/rsos.220411}
\BIBentrySTDinterwordspacing

\bibitem{Daly2008}
\BIBentryALTinterwordspacing
R.~M. Daly, H.~G. Ahlborg, K.~Ringsberg, P.~Gardsell, I.~Sernbo, and M.~K.
  Karlsson, ``{Association Between Changes in Habitual Physical Activity and
  Changes in Bone Density, Muscle Strength, and Functional Performance in
  Elderly Men and Women},'' \emph{Journal of the American Geriatrics Society},
  vol.~56, no.~12, pp. 2252--2260, 2008. [Online]. Available:
  \url{https://doi.org/10.1111/j.1532-5415.2008.02039.x}
\BIBentrySTDinterwordspacing

\bibitem{Morena2021}
\BIBentryALTinterwordspacing
J.~M. Delfa-de~la Morena, E.~A. Castro, M.~A. Rojo-Tirado, and D.~Bores-Garcia,
  ``{Relation of Physical Activity Level to Postural Balance in Obese and
  Overweight Spanish Adult Males: A Cross-Sectional Study},''
  \emph{International Journal of Environmental Research and Public Health},
  vol.~18, no.~16, 2021. [Online]. Available:
  \url{https://doi.org/10.3390/ijerph18168282}
\BIBentrySTDinterwordspacing

\bibitem{Rhonda2018}
\BIBentryALTinterwordspacing
R.~Cohen, B.~Baluch, and L.~J. Duffy, ``{Personality differences amongst drag
  racers and archers: implications for sport injury rehabilitation},''
  \emph{Journal of Exercise Rehabilitation}, vol.~14, no.~5, pp. 783--790,
  2018. [Online]. Available: \url{https://doi.org/10.12965/jer.1836350.175}
\BIBentrySTDinterwordspacing

\bibitem{Mikicin2022}
\BIBentryALTinterwordspacing
M.~Mikicin, ``{Relationships of attention and arousal are responsible for
  action in sports},'' \emph{Biomedical Human Kinetics}, vol.~14, no.~1, pp.
  229--235, 2022. [Online]. Available:
  \url{https://doi.org/10.2478/bhk-2022-0028}
\BIBentrySTDinterwordspacing

\bibitem{Koch2021}
\BIBentryALTinterwordspacing
P.~Koch and B.~Krenn, ``{Executive functions in elite athletes - Comparing
  open-skill and closed-skill sports and considering the role of athletes' past
  involvement in both sport categories},'' \emph{Psychology of Sport and
  Exercise}, vol.~55, p. 101925, 2021. [Online]. Available:
  \url{https://doi.org/10.1016/j.psychsport.2021.101925}
\BIBentrySTDinterwordspacing

\bibitem{Piepiora2023}
\BIBentryALTinterwordspacing
P.~Piepiora and A.~Naczyńska, ``{Personality Traits vs. Sports Classes of
  Polish Representatives in Junior Sports Acrobatics},'' \emph{Sports},
  vol.~11, no.~10, 2023. [Online]. Available:
  \url{https://doi.org/10.3390/sports11100194}
\BIBentrySTDinterwordspacing

\bibitem{Adamczyk2023}
\BIBentryALTinterwordspacing
J.~G. Adamczyk, ``{Support Your Recovery Needs (SYRN) - a systemic approach to
  improve sport performance},'' \emph{Biomedical Human Kinetics}, vol.~15,
  no.~1, pp. 269--279, 2023. [Online]. Available:
  \url{https://doi.org/10.2478/bhk-2023-0033}
\BIBentrySTDinterwordspacing

\bibitem{Ingwersen2023}
C.~K. Ingwersen, C.~Mikkelstrup, J.~N. Jensen, M.~R. Hannemose, and A.~B. Dahl,
  ``{SportsPose: A Dynamic 3D Sports Pose Dataset},'' in \emph{Proceedings of
  the IEEE/CVF International Workshop on Computer Vision in Sports}, 2023.

\bibitem{PachavaGoel2022}
\BIBentryALTinterwordspacing
S.~R. Pachava and M.~Goel, ``{Effect of soft tissue manipulation of popliteus
  muscle on quadriceps muscle activity and torque production in athletes with
  anterior knee pain},'' \emph{Biomedical Human Kinetics}, vol.~14, no.~1, pp.
  102--108, 2022. [Online]. Available:
  \url{https://doi.org/10.2478/bhk-2022-0013}
\BIBentrySTDinterwordspacing

\bibitem{Dupuy2024}
\BIBentryALTinterwordspacing
A.~Dupuy, M.~J. Campbell, H.~A. J., and A.~J. Toth, ``{On the necessity for
  biomechanics research in esports},'' \emph{Sports Biomechanics}, vol.~0,
  no.~0, pp. 1--13, 2024. [Online]. Available:
  \url{https://doi.org/10.1080/14763141.2024.2354440}
\BIBentrySTDinterwordspacing

\bibitem{Smerdov2022AISensors}
\BIBentryALTinterwordspacing
A.~Smerdov, A.~Somov, E.~Burnaev, and A.~Stepanov, ``{AI-enabled prediction of
  video game player performance using the data from heterogeneous sensors},''
  \emph{Multimedia Tools and Applications}, 08 2022. [Online]. Available:
  \url{https://doi.org/10.1007/s11042-022-13464-0}
\BIBentrySTDinterwordspacing

\bibitem{Smerdov2019Chair}
A.~Smerdov, E.~Burnaev, and A.~Somov, ``{eSports Pro-Players Behavior During
  the Game Events: Statistical Analysis of Data Obtained Using the Smart
  Chair},'' in \emph{{2019 IEEE SmartWorld, Ubiquitous Intelligence \&
  Computing, Advanced \& Trusted Computing, Scalable Computing \&
  Communications, Cloud \& Big Data Computing, Internet of People and Smart
  City Innovation}}, 2019, pp. 1768--1775.

\bibitem{Smerdov2021Burnout}
\BIBentryALTinterwordspacing
A.~Smerdov, A.~Somov, B.~Burnaev, Evgeny nd~Zhou, and P.~Lukowicz, ``{Detecting
  Video Game Player Burnout With the Use of Sensor Data and Machine
  Learning},'' \emph{IEEE Internet of Things Journal}, vol.~8, no.~22, pp.
  16\,680--16\,691, 2021. [Online]. Available:
  \url{https://doi.org/10.1109/JIOT.2021.3074740}
\BIBentrySTDinterwordspacing

\bibitem{Nicholson2024EnergyExpenditure}
\BIBentryALTinterwordspacing
M.~Nicholson, D.~Poulus, R.~Robergs, V.~Kelly, and C.~McNulty, ``{How Much
  Energy Do E'Athletes Use during Gameplay? Quantifying Energy Expenditure and
  Heart Rate Variability Within E'Athletes},'' \emph{Sports Medicine - Open},
  vol.~10, no.~1, p.~44, 04 2024. [Online]. Available:
  \url{https://doi.org/10.1186/s40798-024-00708-6}
\BIBentrySTDinterwordspacing

\bibitem{Katelhut2024}
\BIBentryALTinterwordspacing
S.~Ketelhut and C.~R. Nigg, ``{Heartbeats and high scores: esports triggers
  cardiovascular and autonomic stress response},'' \emph{Frontiers in Sports
  and Active Living}, vol.~6, 2024. [Online]. Available:
  \url{https://www.frontiersin.org/articles/10.3389/fspor.2024.1380903}
\BIBentrySTDinterwordspacing

\bibitem{Lam2022}
\BIBentryALTinterwordspacing
W.-K. Lam, B.~Chen, R.-T. Liu, J.~C.-W. Cheung, and D.~W.-C. Wong, ``{Spine
  Posture, Mobility, and Stability of Top Mobile Esports Athletes: A Case
  Series},'' \emph{Biology}, vol.~11, no.~5, 2022. [Online]. Available:
  \url{https://www.mdpi.com/2079-7737/11/5/737}
\BIBentrySTDinterwordspacing

\bibitem{Staniszewski2016}
\BIBentryALTinterwordspacing
M.~Staniszewski, P.~Zybko, and I.~Wiszomirska, ``{Influence of a nine-day
  alpine ski training programme on the postural stability of people with
  different levels of skills},'' \emph{Biomedical Human Kinetics}, vol.~8,
  no.~1, pp. 24--31, 2016. [Online]. Available:
  \url{https://doi.org/10.1515/bhk-2016-0004}
\BIBentrySTDinterwordspacing

\bibitem{Staniszewski2017}
\BIBentryALTinterwordspacing
------, ``{Evaluation of Changes in the Parameters of Body Stability in the
  Participants of a Nine-Day Snowboarding Course},'' \emph{Polish Journal of
  Sport and Tourism}, vol.~24, no.~2, pp. 97--101, 2017. [Online]. Available:
  \url{https://doi.org/10.1515/pjst-2017-0010}
\BIBentrySTDinterwordspacing

\bibitem{Schorderet2021}
\BIBentryALTinterwordspacing
C.~Schorderet, R.~Hilfiker, and L.~Allet, ``{The role of the dominant leg while
  assessing balance performance. A systematic review and meta-analysis},''
  \emph{Gait \& Posture}, vol.~84, pp. 66--78, 2021. [Online]. Available:
  \url{https://doi.org/10.1016/j.gaitpost.2020.11.008}
\BIBentrySTDinterwordspacing

\bibitem{Piropo2021}
U.~Piropo, S.~Costa, I.~Ribeiro, I.~Vidal~Freire, L.~Schettino, R.~Passos,
  M.~Machado, C.~Casotti, and R.~Pereira, ``{Influence of Physically Active or
  Sedentary Lifestyle on Postural Control of Community-dwelling Old Adults},''
  \emph{Exercise Medicine}, vol.~5, p.~1, 12 2021.

\bibitem{Charalambous2019}
\BIBentryALTinterwordspacing
L.~H. Charalambous, R.~B. Champion, L.~R. Smith, A.~C.~S. Mitchell, and D.~P.
  Bailey, ``{Effects of Interrupting Sitting with Use of a Treadmill Desk
  Versus Prolonged Sitting on Postural Stability},'' \emph{International
  Journal of Sports Medicine}, vol.~40, no.~13, pp. 871--875, 12 2019.
  [Online]. Available: \url{https://doi.org/10.1055/a-0975-9313}
\BIBentrySTDinterwordspacing

\bibitem{Dana2021}
\BIBentryALTinterwordspacing
A.~Dana, A.~H. Sabzi, S.~Ghorbani, and A.~G. Rad, ``{The effect of diurnal
  rhythms on static and dynamic balance performance},'' \emph{Biomedical Human
  Kinetics}, vol.~13, no.~1, pp. 205--211, 2021. [Online]. Available:
  \url{https://doi.org/10.2478/bhk-2021-0025}
\BIBentrySTDinterwordspacing

\bibitem{Filipowicz2020}
J.~Filipowicz-Ciepły, J.~Golec, E.~Szczygieł, D.~Czechowska, M.~Cichoń, and
  J.~Balicka-Bom, ``{Estimation of selected parameters in stabilographic image
  depending on physical activity level in children and adolescents},''
  \emph{Medical Rehabilitation}, vol.~24, pp. 4--14, 11 2020.

\bibitem{Ketelhut2023Fitness}
\BIBentryALTinterwordspacing
S.~Ketelhut, A.~Bodman, T.~Ries, and C.~R. Nigg, ``{Challenging the Portrait of
  the Unhealthy Gamer---The Fitness and Health Status of Esports Players and
  Their Peers: Comparative Cross-Sectional Study},'' \emph{J Med Internet Res},
  vol.~25, p. e45063, 08 2023. [Online]. Available:
  \url{https://doi.org/10.2196/45063}
\BIBentrySTDinterwordspacing

\bibitem{Darwesh2023}
\BIBentryALTinterwordspacing
A.~Darwesh, S.~B.~A. Almalhan, S.~B.~M. Bahmeshan, A.~B.~A. Alghamdi, M.~B.~H.
  Alamari, and W.~Qurban, ``{Impact of Prolonged use of Video Gaming on Grip
  and Pinch Strength in Young Adult},'' \emph{Global Journal of Human-Social
  Science}, vol.~23, no.~H8, pp. 29--41, 12 2023. [Online]. Available:
  \url{https://doi.org/10.34257/GJHSSHVOL23IS8PG29}
\BIBentrySTDinterwordspacing

\bibitem{Uluagac2023}
\BIBentryALTinterwordspacing
K.~Ulua{\u{g}}a{\c{c}}, A.~Yeral, and E.~T. {\c{C}}il, ``{The Effects of
  Esports Experience on Hand Function, Strength, Coordination and Pain in Elite
  League of Legend Players: Research Article},'' \emph{Acta Medica Ruha},
  vol.~1, no.~3, pp. 304--315, 09 2023. [Online]. Available:
  \url{https://doi.org/10.5281/zenodo.8252084}
\BIBentrySTDinterwordspacing

\bibitem{Lee2024}
\BIBentryALTinterwordspacing
H.~Lee, S.~Lee, R.~Nallapati, Y.~Uh, and B.~Lee, ``{Characterizing and
  Quantifying Expert Input Behavior in League of Legends},'' in
  \emph{Proceedings of the CHI Conference on Human Factors in Computing
  Systems}, ser. CHI '24.\hskip 1em plus 0.5em minus 0.4em\relax New York, NY,
  USA: Association for Computing Machinery, 2024. [Online]. Available:
  \url{https://doi.org/10.1145/3613904.3642588}
\BIBentrySTDinterwordspacing

\end{thebibliography}

\end{document}